\documentclass[aps,twocolumn,superscriptaddress,floats,floatfix]{revtex4}  
\usepackage{dcolumn}   
\usepackage{amsmath}
\usepackage{epsfig}
\usepackage{epstopdf}
\usepackage{graphicx}
\usepackage{color}
\usepackage{bm}        
\usepackage{amssymb}   
\usepackage{hyperref}
\usepackage[normalem]{ulem}
\usepackage[english]{babel}

\begin{document}

\title{Multivalent ion effects on electrostatic stability of virus-like nano-shells}

\author{Leili \surname{Javidpour}}
\affiliation{School of Physics, Institute for Research in Fundamental Sciences (IPM), Tehran 19395-5531, Iran}
\author{An\v{z}e \surname{Lo\v{s}dorfer Bo\v{z}i\v{c}}}
\affiliation{Department of Theoretical Physics, Jo\v zef Stefan Institute, SI-1000 Ljubljana, Slovenia}
\author{Ali \surname{Naji}}\email{a.naji@ipm.ir}
\affiliation{School of Physics, Institute for Research in Fundamental Sciences (IPM), Tehran 19395-5531, Iran}
\affiliation{Department of Applied Mathematics and Theoretical Physics, University of Cambridge, Centre for Mathematical Sciences, Cambridge CB3 0WA, United Kingdom}
\author{Rudolf \surname{Podgornik}}
\affiliation{Department of Theoretical Physics, Jo\v zef Stefan Institute, SI-1000 Ljubljana, Slovenia}
\affiliation{Department of Physics, Faculty of Mathematics and Physics, University of Ljubljana, SI-1000 Ljubljana, Slovenia}
\affiliation{Department of Physics, University of Massachusetts, Amherst, MA 01003, USA}


\begin{abstract}
Electrostatic properties and stability of charged virus-like nano-shells are examined in ionic solutions with monovalent and multivalent ions. A theoretical model based on a thin charged spherical shell and multivalent ions within the ``dressed multivalent ion'' approximation, yielding their distribution across the shell and the corresponding electrostatic (osmotic) pressure acting on the shell, is compared with extensive implicit Monte-Carlo simulations. It is found to be accurate for positive or low negative surface charge densities of the shell and for sufficiently high (low) monovalent (multivalent) salt concentrations. Phase diagrams involving electrostatic pressure exhibit positive and negative values, corresponding to an outward and an inward facing force on the shell, respectively. This provides an explanation for the high sensitivity of viral shell stability and self-assembly of viral capsid shells on the ionic environment.
\end{abstract}


\maketitle

\section{Introduction}

Aqueous solutions of most macromolecules and macromolecular assemblies with dissociable molecular moieties often exhibit strongly charged surfaces when immersed into a bathing electrolyte (ionic) solution. The electrostatic interaction between macromolecular surface and solution charges in aqueous electrolytes plays a dominant role in determining the thermodynamics properties of charged macromolecular systems, and has been traditionally described using mean-field theories such as Poisson-Boltzmann (PB) or, its linearized version, the Debye-H\"uckel (DH) theory \cite{Israelachvili,andelman-rev}.  In many examples that have  recently received a renewed attention, macromolecules are found in highly asymmetric ionic mixtures containing symmetric monovalent  salts together with asymmetric multivalent salts that contribute mobile ions of large charge valency to the solution. Whereas the electrostatics of systems containing monovalent ions is explained reasonably within the PB (or DH) framework,  multivalent ions are known, from a wealth of both theoretical and experimental results obtained over the last decade,  to cause large deviations from the standard mean-field predictions; these deviations stem from electrostatic correlation and fluctuation effects that are  highly enhanced in the presence of multivalent ions and are deemed
to underly the mechanism behind the like-charge attraction phenomena (see Refs. \cite{Shklv,Levin,holm,hoda,Naji_PhysicaA} and references therein).  The latter are most clearly manifested  in several interesting instances, viz. formation of large aggregates of like-charged polymers such as microtubules \cite{Needleman} and F-actin \cite{Wong}, with the most prominent example being the electrostatic correlation effects in the formation of large condensates of DNA in the bulk \cite{Bloomfield,Yoshikawa1,Yoshikawa2,Pelta} and in the DNA packaging inside viral shells that are observed only in the presence of multivalent cations \cite{Plum,Raspaud,Savithri1987,deFrutos2005,Siber}.

Another example which we shall focus upon in this work is the stability of charged viral capsids and virus-like nano-shells, in particular the contribution of electrostatic interactions to the total free energy of these structures~\cite{PCCP2012}. Bacteriophage capsids have been known to undergo an {\em osmotic shock}, i.e. a shape destabilization and eventual rupture when immersed into distilled water \cite{Pollard,Gelbartreview}. In the osmotic shock the large positive osmotic pressure (typically $\sim 50-100 ~\rm atm$) within the virus capsid, engendered by strong intermolecular repulsions between DNA molecules \cite{Gelbart,Dragar}, causes the regions of the icosahedral capsid between the five-fold symmetric sites to {\em buckle outwards} in order to allow for the volume increase of the capsid driven by the internal pressure \cite{Siberinst}. In the limiting case the capsid is ruptured and the encapsidated DNA spills out, as shown in the bacteriophage T2 iconic image of Kleinschmidt et al. \cite{Kleinschmidt}. Contrary to the action of large positive osmotic pressures, negative osmotic pressure can engender a different type of empty capsid destabilization based on an {\em inward buckling} of the elastic shell \cite{Siber2009}. For a typical virus of $\sim 30 ~\rm nm$ this happens at the critical pressure of about $\sim 5 ~\rm atm$ when the initially icosahedral capsid crumples, showing a complicated structure compatible with elastic soft mode displacement patterns.
It is usually thought that the negative osmotic pressure can be due either to the osmotic stress of the external solution containing an osmoticant, such as PEG (poly-ethylene-oxide), that is excluded from entering the capsid \cite{Evilev}, or by the negative osmotic pressure due to the polyelectrolyte bridging interaction \cite{bridge} of the flexible ss-RNA genome of the virus \cite{RNAosmotic}. In both  cases the destabilizing negative osmotic pressure can easily reach values of a few atmospheres. In the case of the ss-RNA viruses small negative osmotic pressures are in fact needed in order to stabilize the filled virus containing the ss-RNA genome,  with typical values of $\sim - 0.5 ~\rm atm$ at physiological salt conditions, tending towards zero at the border of feasibility of spontaneous self-assembly of filled capsids.

Stability of viruses does not only depend on their genomic cargo generated osmotic pressure or osmotic stress provided by the osmoticants in the external aqueous compartment as elucidated above, but also on the detailed ionic composition of the bathing medium. In their role of stabilizing/destabilizing agents the multivalent ions of the bathing medium behave rather differently from monovalent ions. Some of the major functions of multivalent metal ions such as $\mathrm{Mg}^{2+}$, $\mathrm{Zn}^{2+}$, and $\mathrm{Ca}^{2+}$ are surmised to enhance the capsid stability, enable their conformational alterations~\cite{Rossman}, assist in the genome packaging~\cite{Carrivain2012} while their removal can induce swelling of capsids \cite{Rossman2} or even structural transitions in the capsid \cite{Sherman}. Multivalent ions such as lanthanides $\mathrm{Gd}^{3+}$ or  $\mathrm{Tb}^{3+}$ can be used to stabilize virus-like nano-particles \cite{Steinmetz2011}, while polyamines like spermine $\mathrm{Spm}^{4+}$ are required for proper stabilization of different viruses~\cite{Tyms1989,Cohen1979}, most notably the Belladonna mottle virus \cite{Savithri1987} as well as the tobacco mosaic virus \cite{Torrigiani}. While obviously of such paramount importance for the stability of viruses, the exact mechanism of electrostatic capsid de/stabilization due to the presence of multivalent bathing solution ions remains largely elusive. In what follows we will attempt to fill in this gap in our understanding of the stabilization of viruses in different solution conditions, by showing convincingly that the electrostatically strongly coupled multivalent counterions are yet another universal agent that can change the osmotic pressure acting on the capsid shell driving it from positive to negative values. This will shed some much needed light on the universal mechanisms by which they can act as stabilizing agents for virus capsid self-assembly.

In order to properly understand the detailed nature of electrostatic interactions involving virus capsids, we investigate the effects of highly charged bathing solution ions on the capsid stability via their spatial distribution and the concurrent (osmotic) pressure. To accomplish this goal, the PB paradigm has to be amended since, as noted above, mean-field theories are valid only for weakly charged systems and monovalent ions~\cite{Netz,hoda,Shklv}. This is achieved by employing an analytical  ``dressed multivalent-ion approach''~\cite{Kanduc2010,Kanduc2011,Kanduc2012} as well as by exhaustive and detailed Monte-Carlo (MC) simulations within a simple model that properly captures the salient features of electrostatic effects on capsid stability. We will show that the electrostatically strongly coupled multivalent ions provide a universal mechanism for negative osmotic pressure generation in closed capsid shells, in many respects akin to the action of the bridging RNA genome or external osmoticants.

The paper is organized as follows: In Sec.~\ref{sec:mm} we will introduce our theoretical model as well as the details of our MC simulations.  In Sec.~\ref{sec:val}, we shall compare the results of the simulations with the theoretical predictions for the density profile of multivalent ions next to the capsid and will thus investigate regions of validity of the theoretical model. In Sec.~\ref{sec:res}, we shall consider the electrostatic pressure acting on the capsid and discuss different situations with negative total electrostatic pressure acting on the capsid and summarize our results in Sec.~\ref{sec:sum}.

\section{Model and methods}
\label{sec:mm}

\begin{figure}[t!]
\includegraphics[width=5cm]{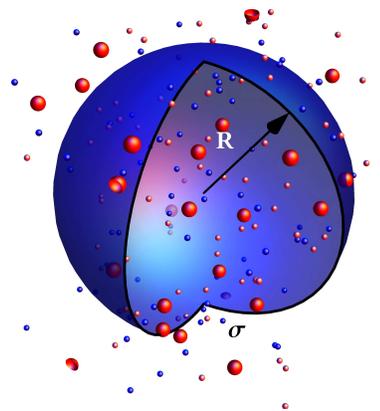}
\caption{Schematic view of a permeable spherical shell of radius $R$ and surface charge density $\sigma$ immersed in a mixed ionic solution comprising monovalent and multivalent salt ions (multivalent ions are shown by bigger red spheres).}
\label{fig:sketch}
\end{figure}

Consider a closed, thin spherical shell of effective radius $R$ and surface charge density $\sigma$, Fig.~\ref{fig:sketch}, which can serve as a reasonable model for a wide range of viral capsids~\cite{Baker1999,ALB2012} as well as synthetic nano-shells~\cite{Steinmetz2011,Yildiz2012}. For typical virus capsids, the average radii range between $5~{\mathrm{nm}}\lesssim R \lesssim30~{\mathrm{nm}}$, with the corresponding thicknesses of the low-dielectric-constant shell material  usually between 1.5 and 4.5~nm. The ensuing finite shell thickness effects are then typically small
at physiological salt conditions where the characteristic Debye length corresponding to the
monovalent salt would be close to 1~nm \cite{Siber2007}. The values of surface charge densities on the capsids are $|\sigma|\lesssim 0.4$ $e_0/\mathrm{nm}^{2}$, and there is no clear-cut preference for inner-outer capsid surface charge density sign segregation, with both signs being admissible~\cite{ALB2012}. The shell is assumed to be permeable to all ions and immersed in an ionic solution comprised of a monovalent $1:1$ salt of bulk concentration $n_0$ and a multivalent $q:1$ salt of bulk concentration $c_0$, with $q$ being the charge valency of multivalent ions.

The above model in its abroad features presents an inhomogeneous, asymmetric, multicomponent Coulomb system which is not amenable to exact analytical treatment. In order to elucidate its behaviour we have to resort to simplifying approximations among which we found the ``dressed multivalent-ion approximation'' to be most helpful. It can be derived from a systematic field-theoretical formalism applicable to highly asymmetric ionic systems with $q>1$~\cite{Kanduc2010,Kanduc2011}. Within this framework, the degrees of freedom associated with monovalent ions are explicitly integrated out, yielding an effective formalism incorporating only the screened Debye-H\"uckel (DH) interactions between the remaining  ``dressed'' multivalent ions and fixed (macroion) surface charges~\cite{Kanduc2010,Kanduc2011}. While this is still a many-body problem, it can be treated with fast and efficient MC simulations involving only dressed ions' degrees of freedom.

On the analytical level, the free energy can be expanded in terms of the fugacity (concentration) of multivalent ions, which is  invariably small in most cases of experimental interest~\cite{Savithri1987,deFrutos2005}. To the leading order this virial expansion leads to the {\em dressed multivalent-ion theory} based on the lowest-order {\em single-particle} contribution from the interaction between dressed multivalent ions and fixed macroion charges. As such it is thus directly related to the {\em strong-coupling theory}~\cite{Netz,hoda,Shklv}, introduced originally for salt-free systems containing only multivalent counterions and describing the limit {\em opposite} to the usual PB or DH theories. This leading-order theory incorporates ion-surface correlations, which are dominant for large $q$ as compared with the subdominant ion-ion contributions. It has been tested extensively with (explicit) MC simulations and shown to provide an accurate description for large $q$, sufficiently large (small) monovalent (multivalent) salt concentrations~\cite{Kanduc2010,Kanduc2011,Kanduc2012}.

For a charged shell, one can express the dressed multivalent-ion free energy of the system as~\cite{Kanduc2010,Kanduc2011,Kanduc2012}
\begin{equation}
\label{eq:freeen}
\beta {\mathcal F}= \beta {\mathcal F}_{DH} - c_0\int\mathrm{d}{\mathbf r} \, e^{-\beta qe_0\varphi_{DH}({\mathbf r})} + {\cal O}(c_0^2),
\end{equation}
where $\beta=1/k_{\mathrm{B}}T$, ${\mathcal F}_{DH}$ is the DH self-energy of the shell in the absence of multivalent ions, and the second term corresponds to the single-particle partition function of a multivalent ion interacting with the shell. Though the above dressed multivalent-ion free energy contains DH components, it is important to note that it differs fundamentally from the standard DH free energy \cite{Netz,hoda} as it follows from the virial expansion in terms of the multivalent-ion fugacity.

The screened DH electrostatic potential of the permeable shell, $\varphi_{DH}({\mathbf r})$, entering Eq. \ref{eq:freeen}, is obtained from the standard DH equation. The explicit forms of $\varphi_{DH}({\mathbf r})$ and ${\mathcal F}_{DH}$ are given in Ref.~\cite{Siber2007}, and are
\begin{equation}
\varphi_{DH}(r\leqslant R)=\frac{\sinh\kappa r}{r}\frac{\sigma R}{\kappa\varepsilon\varepsilon_0(\cosh\kappa R+\sinh\kappa R)},
\end{equation}
\begin{equation}
\varphi_{DH}(r\geqslant R)=\frac{e^{-\kappa(r-R)}}{r}\frac{\sigma R}{\kappa \varepsilon\varepsilon_0(1+\coth\kappa R)},
\end{equation}
and
\begin{equation}
\nonumber {\mathcal F}_{DH}=\frac{2\pi R^2\sigma^2}{\kappa\varepsilon\varepsilon_0(1+\coth\kappa R)}.
\end{equation}
The inverse Debye screening length $\kappa$ of the DH potential depends on the bulk concentrations of both monovalent and multivalent ions as
\begin{equation}
\kappa^2=8\pi \ell_{\mathrm{B}}\left(n_0+\frac{qc_0}{2}\right),
\label{eq:kappa_n_c}
\end{equation}
where $\ell_{\mathrm{B}}=\beta e_0^2/(4\pi\varepsilon\varepsilon_0)$ is the standard Bjerrum length.

\begin{figure*}[t!]
\includegraphics[width=18cm]{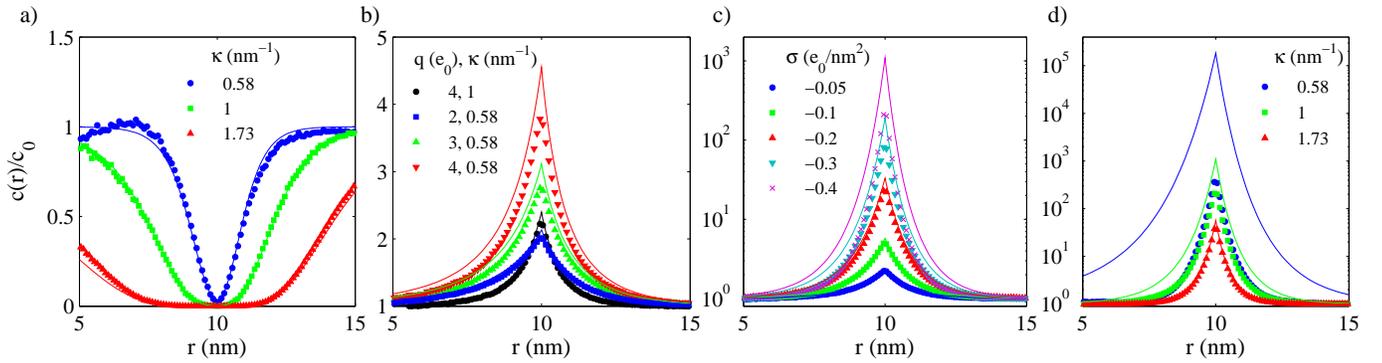}
\caption{(Color online) a) Rescaled radial density  of multivalent ions, $c(r)/c_0$, in the vicinity of a charged shell with radius of $R=10$~nm as a function of the radial distance from the center of the shell. Symbols are MC data and lines analytical predictions of the dressed multivalent-ions theory. Results are shown for a) $\sigma = +0.4$ $e_0/\mathrm{nm}^2$, $q = 4$, and various $\kappa$, b) $\sigma = -0.05$ $e_0/\mathrm{nm}^2$, with $\kappa$ and $q$ values as shown on the graph, c)  $\kappa=1$ ${\mathrm{nm}^{-1}}$, $q = 4$, and different values of (negative) $\sigma$, d) $\sigma = -0.4\, e_0/\mathrm{nm}^2$,  $q = 4$, and different  $\kappa$. In all cases, we have kept $q c_0 = 4$~mM constant.}
\label{fig:Profiles}
\end{figure*}

Furthemore, the (number) density of dressed multivalent ions follows as~\cite{Kanduc2010,Kanduc2011,Kanduc2012}
\begin{equation}
c({\mathbf r}) = c_0 \,\exp\big(-\beta qe_0\varphi_{DH}({\mathbf r})\big),
\label{eq:dens}
\end{equation}
and the electrostatic (osmotic) pressure acting on the shell,  $P= - \partial {\mathcal F}/{\partial V}$,  is given by
\begin{equation}
P = -\frac{\partial {\mathcal F}_{DH} }{\partial V} + k_{\mathrm{B}}T c_0\, \frac{\partial} {\partial V}\!\!\int\!\mathrm{d}{\mathbf r} \, e^{-\beta qe_0\varphi_{DH}({\mathbf r})}\equiv P_{DH} + P_q,
\label{eq:pressure}
\end{equation}
where $V=4\pi R^3/3$ is the shell volume, and the {\em total charge} ${Q_\sigma}=4\pi \sigma R^2$ is kept fixed. The pressure can be divided into two parts: the purely repulsive (positive) $P_{DH}$ from the self-energy of the charged shell, and $P_q$ which results from the multivalent ions and can be attractive (negative), as we show later. It is this last part of the osmotic pressure that will play a fundamental role in elucidating the stabilizing action of multivalent cations on virus caspids.

\subsection{MC simulations}

In order to assess the validity of the  theory detailed above we have used extensive MC simulations in which we simulate a large collection of mobile multivalent ions near a charged permeable shell, with the monovalent ion component treated implicitly, providing a screened DH potential for the multivalent component. In the simulations the capsid is modeled as a thin permeable spherical shell and the ions are assumed to be point-like (in practice, we assume a repulsive shifted Lennard-Jones potential between multivalent ions with an effective diameter of 1\AA~but the
multivalent ions remain well apart due to their Coulomb repulsion and the details of the Lennard-Jones potential do not matter).
The monovalent salt concentration, $n_0$, is varied in the range between 30~mM to  300~mM and we typically take a small multivalent salt concentration $c_0$ of a few mM (e.g., around 1-3 mM of tetravalent ions), in accordance with the typical values found in experiments~\cite{Savithri1987,deFrutos2005}. In what follows, we consider {\em positively} charged multivalent ions with $q=2$, 3, 4, the temperature of $T=300$~K, and dielectric constant of water, $\varepsilon=80$. The inverse Debye screening length typically changes between $\kappa = 0.58-1.73\, {\mathrm{nm}^{-1}}$ (note that for the given range of parameter the contribution of multivalent ions to the Debye screening length is quite small as compared with the monovalent ions, see Eq. (\ref{eq:kappa_n_c})).

We use canonical MC simulations for dressed multivalent ions confined to a sufficiently large outer cubic volume $V$ with the charged shell located at the center. The size of the outer cubic box is taken as $2(R+12\kappa^{-1})$ and we implement periodic boundary conditions in order to minimize the effects of the outer confinement, which upon further inspection is found to be unimportant and does not affect the simulation results within the regime of parameters under consideration here. The simulations typically run for $(6-10)\times10^9$ MC steps with $(1-5)\times10^9$ steps being used for equilibration and the rest for collecting data in order to evaluate thermodynamic averages. It is to be noted that the number of particles (multivalent ions) used in the simulations varies depending on the systems parameters 
and may range from a few tens up to a few hundreds  of particles in the simulations box (for instance, for tetravalent ions of bulk concentration $c_0=1$~mM, we have 43 particles for  $\sigma=-0.4 \, e_0/\mathrm{nm}^2$, $\kappa=1 \, {\mathrm{nm}^{-1}}$, and $R=5$~nm and 802 particles for $\sigma=-0.4 \, e_0/\mathrm{nm}^2$, $\kappa=0.58 \, {\mathrm{nm}^{-1}}$, $R=20$~nm).
The number of ions in our simulations is not given {\em a priori} as an input parameter but is determined self-consistently and in such way that it yields the desired value of the bulk concentration of
multivalent ions, $c_0$ (being defined as the constant value given by the plateau region of the density profile of ions at far distances from the capsid) for any given set of parameters. For this
purpose we employ an iterative method by starting with an initial guess for the number of particles in the box (which can be taken,
for instance, as $c_0V$ or $c_0V$ plus/minus the number of ions that are necessary to give a total
charge equivalent to the negative/positive charge of the nano-container). We then simulate this system and obtain the resultant (equilibrium) bulk concentration. This outcome   
may in general differ from $c_0$, but it can be used to get another estimate for the number of particles that, upon running the simulations again, can generate 
a closer value to the desired  bulk concentration. 
This procedure converges iteratively and it is thus repeated until the bulk value $c_0$ is achieved within an accuracy of 2\%.	

The electrostatic pressure acting on the capsid can be obtained from the MC simulations using the following relation:
\begin{equation}
P=P_{DH}-\left\langle\sum_i q_i\frac{\partial \varphi_{DH}({\bf r}_i)}{\partial V}\Big|_{Q_\sigma}\right\rangle.
\label{nuisgeorni}
\end{equation}
The first term, $P_{DH}$, is the same as in our theoretical model discussed above, and the second correlation contribution contains the effects of multivalent ions. It is obtained from averaging over a sufficiently large MC sample obtained from the simulations after proper equilibration. Note that this quantity, as calculated within the simulations,  in general contains {\em all many-body contributions} from the interactions between multivalent ions and can thus be different from the approximate {\em single-particle} expression $P_q$  used with the dressed multivalent ion theory. In order to calculate this contribution in the simulations, one requires the following expressions, i.e.
\begin{equation}
\frac{\partial\varphi_{DH}({\bf r}_i)}{\partial V}\Big|_{Q_\sigma}=-\frac{\sigma e^{-\kappa R}\sinh\kappa r}{4\pi R^2\varepsilon\varepsilon_0\kappa r}(1+\kappa R),
\end{equation}
for $r\leqslant R$ and
\begin{equation}
\frac{\partial\varphi_{DH}({\bf r}_i)}{\partial V}\Big|_{Q_\sigma}=-\frac{\sigma e^{-\kappa(r-R)}(1-\kappa R\coth\kappa R)}{4\pi R^2\varepsilon\varepsilon_0\kappa r (1+\coth\kappa R)}
\end{equation}
 for $r\geqslant R$. The above two expressions inserted into Eq. (\ref{nuisgeorni}) and averaged over the MC ensemble then give the total osmotic pressure.


\section{Distribution of multivalent ions}
\label{sec:val}

In Fig.~\ref{fig:Profiles}, we show the rescaled number density of multivalent ions as a function of the radial distance $r$ from the center of the shell with $R=10$~nm. In certain regimes of parameters, the analytical prediction of Eq.~(\ref{eq:dens}) (solid lines) closely matches the MC results (symbols). For positively charged shells, there is always an excellent agreement, with the number density of ions exhibiting a drop close to the shell surface (Fig.~\ref{fig:Profiles}a). For negatively charged shells, multivalent ions are attracted to the shell and accumulate more strongly at its surface. There is still a wide range of parameters where the approximate dressed multivalent-ion theory and MC simulations quantitatively agree (Figs.~\ref{fig:Profiles}b and c, e.g., for divalent ions and $\kappa=0.58$ $\mathrm{nm}^{-1}$ which is obtained by using 30~mM of 1:1 salt and 2~mM of 2:1 salt). However, by increasing $q$ (Fig.~\ref{fig:Profiles}b), that leads to an increase of the multivalent ion density at the surface, the deviations between the simulations and the analytical predictions are observed to grow, too. This implies an incipient role for multi-particle correlations which are absent in the effective single part ice approximate theory. In all cases where there is a deviation, the theory appears to overestimate the density of ions in the vicinity of the shell. For large $q$, an improved agreement can be achieved by increasing $\kappa$ which reduces multi-particle interactions; e.g., for $q = 4$ the analytical prediction works very well for $\kappa=1$ $\mathrm{nm}^{-1}$, which is obtained by using 93~mM of 1:1 salt and 1~mM of 4:1 salt (black circles).

A similar trend is observed when varying the value of the negative surface charge density. In this case, a large increase in surface density of tetravalent ions follows the modest increase in $\sigma$ from $-0.05$ $e_0/\mathrm{nm}^2$ to $-0.4$ $e_0/\mathrm{nm}^2$ (Fig.~\ref{fig:Profiles}c). The theory again works for sufficiently low $|\sigma|$ and breaks down at larger values. Note that the deviation between the theory and simulations can be tuned further by changing the screening $\kappa$: the validity of the theory for $\sigma = -0.4$ $e_0/\mathrm{nm}^2$ is restored (Fig.~\ref{fig:Profiles}d) if the screening parameter is increased to $\kappa=1.73$ $\mathrm{nm}^{-1}$, which can be achieved by using 282~mM of 1:1 salt and 1~mM of 4:1 salt. The breakdown of the theory is more dramatically observed at $\kappa=0.58$ $\mathrm{nm}^{-1}$ with both limits falling well within the experimentally relevant values. Thus, the applicability of the above analytical predictions depends on all three parameters, $q$, $\sigma$ and $\kappa$.

\begin{figure}[t!]
\includegraphics[width=8.5cm]{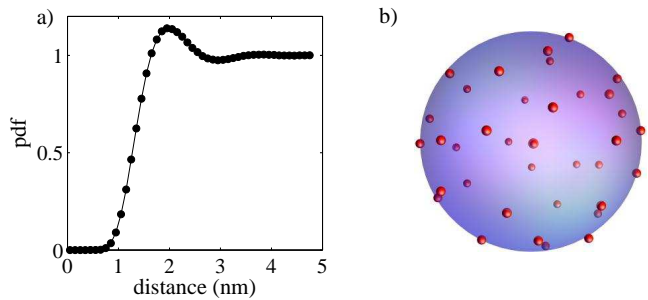}
\caption{(Color online) a) Normalized pair distribution function of multivalent ions from MC simulations for $R = 2.5$~nm, $\sigma = -0.8\, e_0/\mathrm{nm}^2$, $\kappa = 1\,{\mathrm{nm}}^{-1}$, $q=4$. b) A typical snapshot showing multivalent ions in the vicinity (within 1\AA) of the shell for $R = 2.5$ nm, $\sigma = -1.6\, e_0/\mathrm{nm}^2$, $\kappa = 1.73$ ${\mathrm{nm}}^{-1}$,  $q=4$. For these parameters, the shell experiences a total negative  pressure of (a) $-0.28~k_{\mathrm{B}}T/\mathrm{nm}^3$ and (b) $-1.31~k_{\mathrm{B}}T/\mathrm{nm}^3$.}
\label{fig:pairdist}
\end{figure}

While the density of the multivalent ions {\em across} the shell surface can be understood within the phenomenology of the dressed multivalent-ion theory, the behavior in the lateral directions {\em along} the shell surface can be less so. The pair distribution function (pdf) of multivalent ions along the surface (Fig.~\ref{fig:pairdist}a) shows a noticeable {\em correlation hole} of size $a_\bot\sim \sqrt{qe_0/|\sigma|}$ when plotted in terms of their lateral distance, resulting from the repulsion between ions residing close to the shell and local electroneutrality~\cite{Netz,hoda,Shklv}. More remarkably, in the regime where the theory breaks down (large $|\sigma|$ or small $\kappa$, Fig.~\ref{fig:Profiles}), we find a short-range ordering of multivalent ions along the surface clearly exhibited by a pronounced correlation peak in their pdf. This behavior illustrates the fact that the failure of the presented theory stems from the subleading ion-ion correlations that gain in importance when the accumulation of ions at the shell surface is enhanced~\cite{Netz,hoda,Kanduc2010,Kanduc2011}.

\section{Electrostatic pressure acting on the capsid}
\label{sec:res}

\begin{figure}[t!]
\includegraphics[width=9cm]{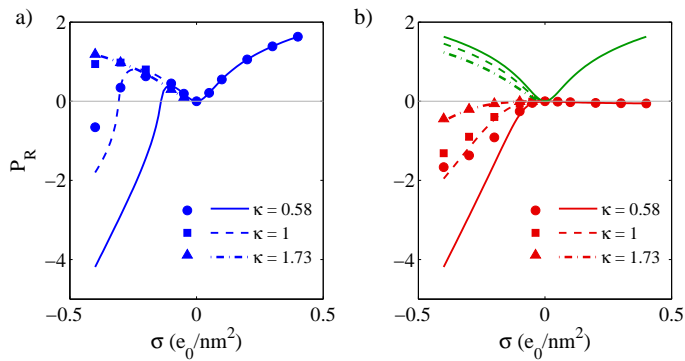}
\caption{(Color online) a) Rescaled total electrostatic pressure $P_{R} = {\rm sgn}(P') \times \log_{10}{(1+100~|P'|)}$ (where $P' = P/P_0$ and $P_0=1$ $k_{\mathrm{B}}T/{\mathrm{nm}}^{3} \simeq 41 ~\rm atm$)
from  simulations (symbols) and the dressed multivalent-ion theory (lines) as a function of $\sigma$ for $R=2.5$ nm, $q = 4$, $c_0=1$~mM, for different values of $\kappa$. b) Same as in panel (a) but showing different components of the total  pressure due to shell's self-energy, $P_{DH}$ (green, upper lines) and multivalent ions, $P_q$ (red symbols and red, lower lines). The rescaling of the pressure is done in such a way that the negative and positive values of pressure can be shown in a log-linear scale and points of zero pressure are correctly represented.
}
\label{fig:pressure}
\end{figure}

The electrostatic pressure acting on the shell also shows interesting variation with system parameters; as an example we vary the surface charge density $\sigma$ (Fig.~\ref{fig:pressure}). For $\sigma>0$, the pressure obtained from simulations is always positive and agrees perfectly with the analytical theory, see Fig.~\ref{fig:pressure}a (blue lines). For $\sigma<0$, both the theory and simulations give {\em negative} pressure for the screening parameter around the physiological value or smaller, indicating that the shell is being {\em compressed}. In this case the theory again overestimates the pressure and shows deviations from the simulations.

When the repulsive self-energy pressure $P_{DH}$ is subtracted, one remains with the correlation pressure imparted on the shell by multivalent ions, $P_q$, shown in red symbols (simulations) and lines (theory), Fig.~\ref{fig:pressure}b. Clearly, the negative pressure thus results purely from multivalent ions and only in the case when the shell is negatively (oppositely) charged. This negative pressure is directly related to the attractive interaction mediated by the multivalent ions between equally charged planar surfaces~\cite{Netz,hoda,Kanduc2010}, i.e., from summing up attractive forces exerted by ions along the outer and inner surfaces of the charged shell that have a component directed towards the inside of the shell. Again, to the leading-order, this effect results from ion-surface correlations captured directly within the single-particle dressed multivalent-ion theory as evidenced by the agreement obtained between the $P_q$ values from the theory and simulations in Fig.~\ref{fig:pressure}b for moderately large $\kappa$; deviations start occurring when ion-ion correlation effects grow.

\begin{figure}[t]
\includegraphics[width=9cm]{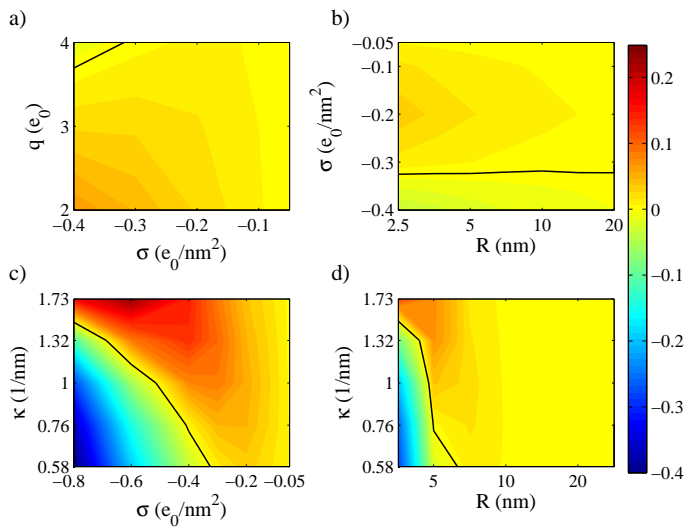}
\caption{(Color online) ``Phase diagrams'' showing lacunae of positive and negative pressure for varying  $q$ and $\sigma$ (shown in linear scale) as well as $R$ and $\kappa$ (shown in logarithmic scale). Solid lines correspond to zero pressure, a) with  $R = 2.5$~nm and $\kappa = 0.58\,{\mathrm{nm}}^{-1}$, b) with $q = 4$ and $\kappa = 0.58\,{\mathrm{nm}}^{-1}$, c)  with $q = 4$ and $R = 2.5$~nm, and d) with $q = 4$ and $Q_\sigma = 4\pi\sigma R^2 = -126~e_0$ (e.g., we have  $\sigma \simeq -0.1$ $e_0/\mathrm{nm}^2$ for $R = 10$ nm). }
\label{fig:phasediag}
\end{figure}

\begin{figure}[t]
\includegraphics[width=7cm]{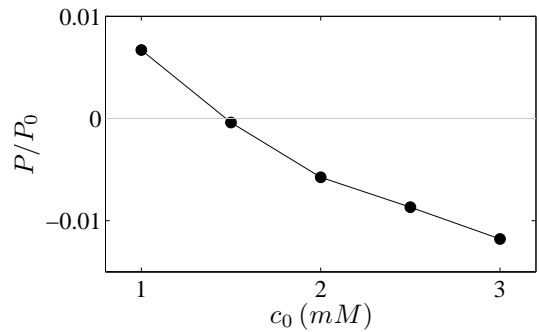}
\caption{Rescaled pressure $P/P_0$ (with $P_0=1\;k_{\mathrm{B}}T/{\mathrm{nm}}^{3} \simeq 41 ~\rm atm$) plotted for different values of multivalent ion concentration $c_0$ for $q=4$, $R=10$ nm, $\sigma =-0.5\,e_0/\mathrm{nm}^2$ and $\kappa = 1\,{\mathrm{nm}}^{-1}$. }
\label{fig:c0}
\end{figure}

In Fig.~\ref{fig:phasediag} we  plot osmotic pressure ``phase diagrams" obtained from our MC simulations for various system parameters $q$, $\sigma$, $\kappa$, and $R$. We show negative and/or positive pressure lacunae, with the black line as {\em separatrix} of zero pressure. It is thus clear that by changing the system parameters one can reach a point where the electrostatic self-repulsion of the capsid is overcome and gives way to an inward force isotropically compressing the capsid with a negative osmotic pressure. This result holds, for instance, in a wide range of capsid radii at physiological conditions ($n_0\sim 100$~mM), provided that $\sigma$ and/or $c_0$ are large enough 
e.g., for $R=10$~nm and $\sigma= -0.5$ $e_0/\mathrm{nm}^{2}$ and $\kappa = 1 \mathrm{nm}^{-1}$ , the pressure changes sign and becomes negative when $c_0$ is increased from 1~mM to 2~mM (see Fig.~\ref{fig:c0}). This shows that the negative osmotic pressure generated by the multivalent ions correlations close to the capsid surface can be realistically expected in real situations where the effective multivalent salt concentrations can be even larger than those assumed here.

\section{Conclusion and discussion}
\label{sec:sum}

In this study we have investigated the electrostatic properties of virus-like nano-shells in a highly asymmetric ionic solution of the type that one usually encounters in experiments \cite{Savithri1987,deFrutos2005}. The capsid is modeled as a thin permeable spherical shell in a salt mixture consisting of a monovalent salt and a small concentration of multivalent salt,
thus accentuating the unusual electrostatic effects due to multivalent ions within a plausible model for a wide range of viral capsid parameters~\cite{Baker1999,ALB2012} as well as synthetic virus-like nano-shells~\cite{Steinmetz2011,Yildiz2012}.

We have been able to show that the effects of multivalent ions (assumed here to be positively charged) in the presence of a positively charged shell can be described accurately using the dressed multivalent-ion theory, a variant of a previously introduced strong coupling theory. For negatively charged capsids, the theory holds again but only for low negative surface charge densities of the shell and for sufficiently high (low) monovalent (multivalent) salt concentrations. Both signs of the surface charge density can in fact be admissible and are actually observed for various viral capsids~\cite{ALB2012}. Our MC simulations provide the means to explore different regimes in the physical parameter space for this system and thus also allow one to determine the regime of applicability of the dressed multivalent-ion theory.


The sign and magnitude of the osmotic pressure acting on the capsid shell are both crucial in assessing the capsid stability towards shrinking and/or swelling. Sufficiently large negative pressure can in fact lead to inward buckling of the capsid through a sequence of elastic instabilities as demonstrated in Ref.~\cite{Siber2009}. For intermediate values of the F\" oppl-von Karman numbers between 300 and 3000 the critical buckling pressure $P_{c}$ scales inversely with the volume of the shell. On the contrary, for smaller or larger F\" oppl-von Karman numbers it scales inversely with the area of the shell. Thus, in the former regime, when the negative pressure is large enough and exceeds a critical value of $P_{c} \sim -5~[\mathrm{atm}]\times (30~[\mathrm{nm}]/R)^3$, the shell would be elastically destabilized and would buckle inwardly. Our results indicate that in the range of $\sigma \sim -0.5$ $e_0/\mathrm{nm}^{2}$ and $c_0\sim 1$~mM, the negative osmotic pressure would be comparable to the critical value of inward buckling destabilization for large shells with radius $R > 150$~nm. This is an upper bound estimate and could be lowered significantly by increasing the concentration of multivalent ions. Our aim here is not that much to track numerical values but to show, that the effect we are describing is realistic even for parameters that are borderline to those observed or set in actual experiments, where the values of the multivalent salt can be much higher than $c_0\sim 1$~mM.

On the other hand, the change of sign of the osmotic pressure on complete removal of the multivalent ions, thus turning the negative pressure to a positive one, would obviously lead to swelling, possibly even completely destabilizing the capsid towards outward buckling~\cite{Rossman2}. Multivalent ions can thus be implicated in both types of instabilities, leading to larger effects for larger capsids. Interestingly, larger viruses such as mimivirus, megavirus, etc., all seem to have either an external or an internal lipid membrane~\cite{Colson2012} that could serve as a local reservoir for multivalent ions changing the osmotic pressure acting on the shell in either direction. In general, local multivalent salt concentration gradients generated by a semipermeable membrane could thus play an important role in regulating the stability of other enveloped viruses.

Negative osmotic pressure has important consequences also for the self-assembly of viral shells. Clearly for $P>0$, the size of the capsid spontaneously increases, precluding a stable state with a finite $R$. On the other hand, for $P<0$ there could be two conceivable outcomes: strong electrostatic coupling would provide effective attractive interactions between equally charged capsomers, thus promoting their aggregation but not necessarily also their self-assembly into a fully formed capsid. A disordered aggregate might be a more stable configuration than an ordered spherical shell, indicating that salt and multivalent ions could interfere with self-assembly. There are indications, mostly for divalent salts, that this is indeed the case and that solution conditions ($n_0$ of salt, $q$ and $c_0$ of multivalent ions) do play an important role in the assembly/disassembly process~\cite{Rossman,Rossman2}.

At the end we note that our model can be improved by extending it to include more molecular details consistent with viral capsids such as the finite capsid radius, the non-trivial charge distribution of the capsids, and the independent variation of the charge on the outer (epitopal) and inner (hypotopal) surface of the capsid shell~\cite{ALB2012}, etc. The effect of finite capsid thickness (typically $\sim 1.5-4.5$~nm~wide~\cite{ALB2012}) is expected to be small compared with the screening length due to realistic salt concentrations~\cite{Siber2007}. The ion size can also become important for small capsid radii, but  our results (e.g., Fig. \ref{fig:phasediag}b and \ref{fig:c0}) show that some of the interesting effects discussed here (such as negative or zero total pressure and the destabilization threshold) persist also in the regime of large capsid sizes, where the role of ion size is obviously less important.

\begin{acknowledgements}
A.L.B. and R.P. acknowledge support from the ARRS Grants No. P1-0055 and J1-4297. A.N. acknowledges support from the Royal Society, the Royal Academy of Engineering, and the British Academy.
\end{acknowledgements}

\end{document}